\documentclass[conference]{IEEEtran}

\IEEEoverridecommandlockouts   

\usepackage[mode=buildmissing]{standalone} 
\usepackage{amsmath}
\usepackage{bm}
\usepackage[nolist]{acronym}
\usepackage{amssymb}
\usepackage{comment}
\usepackage{graphicx}
\usepackage{color,colortbl}
\usepackage{xcolor}

\usepackage[hidelinks]{hyperref}
\usepackage{cleveref}
\usepackage{verbatim}
\usepackage{enumerate}
\usepackage[shortlabels]{enumitem}
\usepackage{cuted, nccmath}
\usepackage{dblfloatfix}
\usepackage{float} 
\usepackage{lipsum}
\usepackage{balance}
\usepackage{cite}
\usepackage{url}
\usepackage{tabularx}
\usepackage{boldline}
\usepackage{balance}    
\usepackage{mathtools} 
\usepackage{stmaryrd} 
\usepackage{optidef}
\usepackage{graphicx}
\usepackage{pgfplots}
\usepackage{booktabs}
\usepackage[font={footnotesize}]{caption}
\usepackage{subfig} 
\usepackage{tikz}
\usepackage{xparse} 
\usepackage{algorithm}
\usepackage{algpseudocode}

\usepackage[all=normal,paragraphs=tight,floats=normal,mathspacing=normal,wordspacing=normal,charwidths=normal,mathdisplays=normal,leading=normal]{savetrees}

\usepackage{makecell}
\usepackage{multirow}
\usepackage{array}
\usepackage{arydshln}
\usetikzlibrary{arrows.meta}
\tikzset{every picture/.style={line width=0.6pt}}
\usetikzlibrary{arrows.meta,
                calc, 
                backgrounds,
                chains,
                fit,
                quotes,
                shapes.geometric,
                positioning,
                intersections,
                spath3,
                spy}

\algdef{SE}[SUBALG]{Indent}{EndIndent}{}{\algorithmicend\ }%
\algtext*{Indent}
\algtext*{EndIndent}

\definecolor{myred}{rgb}{0,0,0}

\newcommand{\abs}[1]{\ensuremath{\left| #1 \right|}}

\newcommand{\espo}[2]{\ensuremath{\mathbb{E}_{#2}\left[ #1 \right]}}
\DeclareMathOperator*{\argmax}{arg\,max}

\newcommand{\norm}[2]{\ensuremath{\left\lVert #1 \right\rVert}_#2}
\newcommand*{\transp}{{\mkern-1.5mu\mathsf{T}}}
\newcommand*{\herm}{{\mathsf{H}}}

\algnewcommand\algorithmicforeach{\textbf{for each}}
\algdef{S}[FOR]{ForEach}[1]{\algorithmicforeach\ #1\ \algorithmicdo}

\begin{acronym}[ACRONYM]
\acro{6G}{6th generation wireless systems}

\end{acronym}

\begin{document}
\bstctlcite{IEEEexample:BSTcontrol}

\title{Physically Parameterized Differentiable MUSIC for DoA Estimation with Uncalibrated Arrays \\
\thanks{The work of Baptiste CHATELIER is supported by the Brittany region. This work is also supported by the Swedish Research Council under VR grant 2022-03007. }
}
\author{
	\IEEEauthorblockN{
        Baptiste Chatelier\IEEEauthorrefmark{3}$^,$\IEEEauthorrefmark{2},
		José Miguel Mateos-Ramos\IEEEauthorrefmark{4}, 
        Vincent Corlay\IEEEauthorrefmark{3},
		Christian H\"{a}ger\IEEEauthorrefmark{4},\\
        Matthieu Crussière\IEEEauthorrefmark{2},
		Henk Wymeersch\IEEEauthorrefmark{4},
		Luc Le Magoarou\IEEEauthorrefmark{2}		}
	\IEEEauthorblockA{
		\IEEEauthorrefmark{3}Mitsubishi Electric R\&D Centre Europe, Rennes, France
	}
    \IEEEauthorblockA{
		\IEEEauthorrefmark{2}Univ Rennes, INSA Rennes, CNRS, IETR-UMR 6164, Rennes, France
	}
    \IEEEauthorblockA{
		\IEEEauthorrefmark{4}Department of Electrical Engineering, Chalmers University of Technology, Sweden
	}
}

\maketitle

\IEEEpeerreviewmaketitle

\begin{abstract}
Direction of arrival (DoA) estimation is a common sensing problem in radar, sonar, audio, and wireless communication systems. It has gained renewed importance with the advent of the integrated sensing and communication paradigm. To fully exploit the potential of such sensing systems, it is crucial to take into account potential hardware impairments that can negatively impact the obtained performance. This study introduces a joint DoA estimation and hardware impairment learning scheme following a model-based approach. Specifically, a differentiable version of the multiple signal classification (MUSIC) algorithm is derived, allowing efficient learning of the considered impairments. The proposed approach supports both supervised and unsupervised learning strategies, showcasing its practical potential. Simulation results indicate that the proposed method successfully learns significant inaccuracies in both antenna locations and complex gains. Additionally, the proposed method outperforms the classical MUSIC algorithm in the DoA estimation task.
\end{abstract}


\section{Introduction}\label{sec:introduction}
Direction of arrival (DoA) estimation refers to the angular estimation of a wavefront impinging on a sensor array. In modern wireless communication systems, DoAs are used for a wide variety of tasks: localization, target tracking, beamforming, or interference management~\cite{pesavento2023three,Corlay24}. The growing interest in integrated sensing and communication (ISAC), where sensing and communication are jointly performed, has further reinforced the need of efficient DoA estimation. Classically, model-based algorithms have been proposed to solve the DoA estimation problem~\cite{Schmidt1986,Barabell83,VanVeen88,Roy1989,Vasylyshyn09}. While the use of subspace methods such as multiple signal classification (MUSIC)~\cite{Schmidt1986}, or estimation of signal parameters via rotational invariance techniques (ESPRIT)~\cite{Roy1989} allows to increase estimation performance, such methods are sensitive to hardware impairments. Several studies have been carried out to assess the impact of hardware impairments on wireless communication performance~\cite{Björnson13,Björnson14,Björnson15,Höhne10 ,chatelier_impact_2022}, but also on sensing performance~\cite{Bozorgi21,Tubail24,Chen24}.

Machine learning has recently emerged as an alternative to classical signal processing methods in many fields of wireless communication~\cite{OShea2017,Wang2017}, including beamforming~\cite{Alkhateeb2018}, channel estimation~\cite{Balevi2020}, localization~\cite{Chatelier2023b} or decoding~\cite{Corlay2022}. However, such ML approaches can be seen as black boxes, leading to minimal interpretability of the learned system. Alternatively, the model-based machine learning paradigm~\cite{Shlezinger23,Shlezinger23b} offers increased interpretability in the learned structure, as models from signal processing are used to structure and initialize neural architectures. This paradigm has been employed in several communication problems: channel estimation~\cite{yassine2022,Chatelier2022efficient}, precoding~\cite{Lavi23}, beam prediction~\cite{Yassine23}, detection~\cite{Samuel17}, and also in ISAC design~\cite{mateos2023model,Mateos_Ramos24}.

\noindent\textbf{Contributions.} It is proposed to tackle the joint DoA estimation and hardware impairments learning problem following the model-based machine learning paradigm. In this paper, the use of a differentiable MUSIC (diffMUSIC) algorithm for antenna array location inaccuracies and complex gains impairments learning is studied. This differentiable version of the classical MUSIC algorithm is achieved by replacing the non-differentiable argmax step by a convex combination of DoAs. Additionally, supervised and unsupervised learning strategies are presented. Experiments conducted on synthetic data against several baselines demonstrate the effectiveness of the proposed method. It is shown that the proposed method is able to compensate for significant antenna array impairments. Specifically, impairments are considered on both the location and complex gain of each radiating elements composing the antenna array.

\noindent\textbf{Related work.} The use of machine learning methods in the DoA estimation context has been studied in the literature: a non-exhaustive list includes~\cite{Wu2019,Cong2021,Papageorgiou2021,Lan2023,Ji2024}. Additionally, the model-based machine learning paradigm has been applied to this task in~\cite{Shmuel2023a, Shmuel2023}. The MUSIC method considers several assumptions on the system model: e.g., sources must be non-coherent, the signals must be narrowband. In~\cite{Shmuel2023a, Shmuel2023}, the authors aim to relax these assumptions by introducing SubspaceNet: a convolutional neural network (CNN) that outputs a surrogate covariance matrix, that can then be used by the MUSIC method. Such method has been shown to counter the effect of coherent sources in addition to hardware impairments. The diffMUSIC method proposed in this paper is aimed at achieving hardware impairments learning without the use of a CNN, with the benefit of drastically reducing the number of learnable parameters.

\section{Problem formulation}\label{sec:problem_formulation}

\begin{figure}
    \centering
    \includegraphics[width=.8\columnwidth]{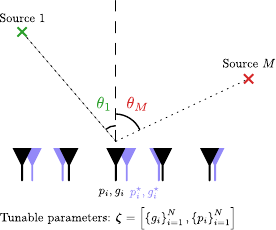}
    \caption{System model: purple antennas are the physical antennas, black antennas represent the nominal antennas}
    \label{fig:system_model}
\end{figure}

A uniform linear array (ULA) composed of $N$ antennas with half-wavelength spacing is considered. The presented approach can be straightforwardly extended to any given antenna array geometry. $M < N$ non-coherent far-field sources are impinging the array. Measurements are carried out over $T$ snapshots:
\begin{equation}\label{eq:system_model}
	\mathbf{X} = \mathbf{A}_{\boldsymbol{\zeta}}\left(\boldsymbol{\theta}\right)\mathbf{S} + \mathbf{N},
\end{equation}
where $\mathbf{X} \in \mathbb{C}^{N \times T}$ corresponds to the measured signals, and $\boldsymbol{\theta} \in \left[-\pi/2,\pi/2\right]^M$ represents the sources DoAs. The narrowband sources signals are defined as $\mathbf{S} = \left\{\mathbf{s}_i\right\}_{i=1}^T \in \mathbb{C}^{M \times T}$, such that $\mathbf{s}_i \sim \mathcal{CN}\left(\mathbf{0}_M,\sigma^2_s \mathbf{I}_M\right)$. The sensing noise $\mathbf{N} = \left\{\mathbf{n}_i\right\}_{i=1}^T \in \mathbb{C}^{N \times T}$ is defined such that $\mathbf{n}_i \sim \mathcal{CN}\left(\mathbf{0}_N,\sigma^2_n\mathbf{I}_N\right)$. $\mathbf{A}_{\boldsymbol{\zeta}}\left(\boldsymbol{\theta}\right) = \left\{\mathbf{a}_{\boldsymbol{\zeta}}\left(\theta_i\right)\right\}_{i=1}^M$ is a steering matrix defined as:
\begin{equation}
	\mathbf{A}_{\boldsymbol{\zeta}}\left(\boldsymbol{\theta}\right) = \left\{\dfrac{1}{\norm{\mathbf{g}}{2}}\begin{bmatrix}
		g_1 \mathrm{e}^{-\mathrm{j}\frac{2\pi}{\lambda} p_1 u\left(\theta_i\right)}\\
		\vdots\\
		g_N \mathrm{e}^{-\mathrm{j}\frac{2\pi}{\lambda} p_N u\left(\theta_i\right)}
	\end{bmatrix}
	\right\}_{i=1}^M \in \mathbb{C}^{N \times M},
\end{equation}
where $\mathbf{g} = \{g_i\}_{i=1}^N \in \mathbb{C}^N$ represents the complex antenna gains, $p_i \in \mathbb{R}$ is the location of the $i$th antenna along the array axis, and $u\left(\theta_i\right) = \cos\left(\theta_i\right)$. Finally, $\boldsymbol{\zeta} = \left[\left\{g_i\right\}_{i=1}^N, \left\{p_i\right\}_{i=1}^N\right]$ corresponds to a physical parametrization of the array manifold. This system model is visualized in Fig.~\ref{fig:system_model}.

The DoA estimation problem can be seen as follows: can the DoAs $\boldsymbol{\theta}$ be recovered from the received signal $\mathbf{X}$? 

\noindent\textbf{Subspace methods.} The eigenstructure of the received signal covariance matrix can be exploited to perform DoA estimation. Indeed, with non-coherent sources, the covariance matrix of the received signal expresses as:
\begin{equation}\label{eq:cov_model}
	\mathbf{\Gamma}_{\mathbf{X}} = \mathbf{A}_{\boldsymbol{\zeta}}\left(\boldsymbol{\theta}\right)\mathbf{\Gamma}_{\mathbf{S}}\mathbf{A}_{\boldsymbol{\zeta}}^\herm\left(\boldsymbol{\theta}\right) + \sigma^2_n \mathbf{I}_N,
\end{equation}
where $\mathbf{\Gamma}_{\mathbf{S}}$, the covariance matrix of $\mathbf{S}$, is diagonal. Since
\begin{equation}
    \mathrm{rank}\left(\mathbf{A}_{\boldsymbol{\zeta}}\left(\boldsymbol{\theta}\right)\mathbf{\Gamma}_{\mathbf{S}}\mathbf{A}_{\boldsymbol{\zeta}}^\herm\left(\boldsymbol{\theta}\right)\right) = M,
\end{equation}
$\mathbf{A}_{\boldsymbol{\zeta}}\left(\boldsymbol{\theta}\right)\mathbf{\Gamma}_{\mathbf{S}}\mathbf{A}_{\boldsymbol{\zeta}}^\herm\left(\boldsymbol{\theta}\right)$ admits only $M$  strictly positive eigenvalues. Furthermore, $\mathbf{\Gamma}_{\mathbf{X}}$ admits the following eigenvalue decomposition (EVD):
\begin{equation}\label{eq:evd}
	\mathbf{\Gamma}_{\mathbf{X}} = \mathbf{U}\mathbf{\Lambda}\mathbf{U}^\herm,
\end{equation}
where $\mathbf{U}$ is the matrix containing the eigenvectors and $\mathbf{\Lambda}$ is the diagonal matrix holding the eigenvalues on its diagonal. Then, combining Eq.~\eqref{eq:cov_model} and~\eqref{eq:evd} yields the following results (see~\cite[Chapter 4, pp.159-164]{Stoica2005} for more details):
\begin{enumerate}
	\item The eigenvalues ordered by decreasing amplitude are $\lambda_1, \cdots, \lambda_M, \lambda_{M+1}, \cdots, \lambda_{M+1}$, where $\lambda_{M+1} = \sigma^2_n$ has multiplicity $N-M$.
	\item The eigenvectors span a space that can be decomposed into signal and noise subspaces: $\mathbf{U} = \left[\mathbf{U}_S, \mathbf{U}_N\right]$, with $\mathbf{U}_S \bot \mathbf{U}_N$.
\end{enumerate}
It is then clearly established that $\mathbf{A}_{\boldsymbol{\zeta}}\left(\boldsymbol{\theta}\right) \bot \mathbf{U}_N$, leading to:
\begin{equation}\label{eq:spectrum}
	\forall i \in \llbracket 1,M\rrbracket, \norm{\mathbf{U}_N^\herm \mathbf{a}_{\boldsymbol{\zeta}}\left(\theta_i\right)}{2}^2=0.
\end{equation}

\begin{figure}[h!]
    \centering
    \includegraphics[width=\columnwidth]{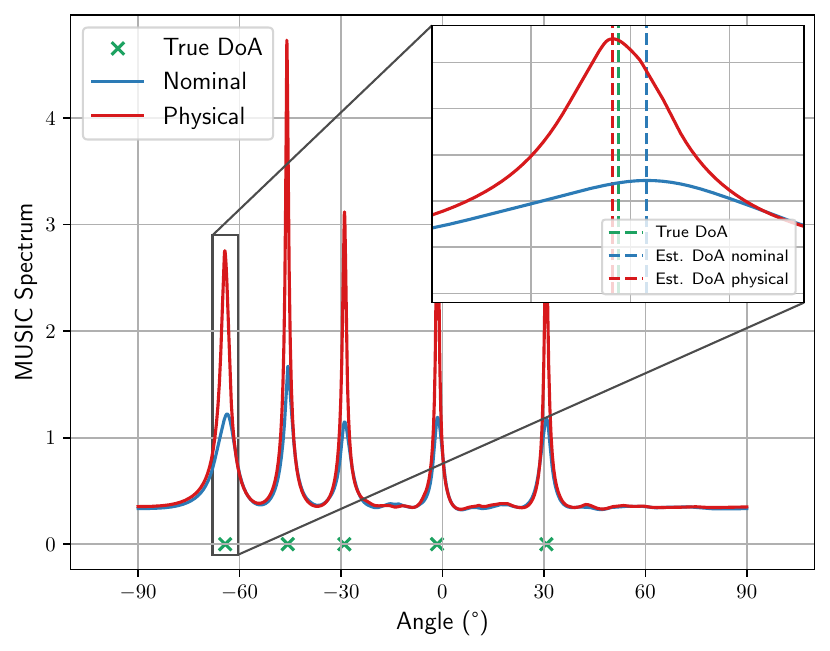}
    \caption{MUSIC performance under gain and location impairments: nominal, resp. physical, represents the spectrum without, resp. with, hardware impairment knowledge}
    \label{fig:MUSIC_hardware_impairments}
\end{figure}

\noindent\textbf{MUSIC.} The core of this method lies in Eq.~\eqref{eq:spectrum} and is summarized in Algorithm~\ref{alg:MUSIC}. Since the inverse of Eq.~\eqref{eq:spectrum} tends to infinity for the sources' DoAs, the MUSIC algorithm involves evaluating this inverse on an angular grid $\boldsymbol{\theta}_g$ and identifying the sources' DoAs as the arguments of the peaks. It is important to note from Eq.~\eqref{eq:argmax_computation} that DoA estimation performance depends on the array parametrization knowledge: imperfect knowledge of the antenna gains and locations leads to error in the MUSIC spectrum peak locations and amplitudes, thereby altering the estimated DoAs, as presented in Fig.~\ref{fig:MUSIC_hardware_impairments}.

\begin{algorithm}[h]
    \caption{MUSIC algorithm.}
    \label{alg:MUSIC}
    \begin{algorithmic}[1]
        \Require Measured signals $\mathbf{X}\in\mathbb{C}^{N\times T}$, number of sources $M$, angular grid $\boldsymbol{\theta}_g \in \mathbb{R}^{N_{\theta}}$, current array parametrization knowledge $\boldsymbol{\zeta}$.
        \State Compute the sample covariance matrix:
        \begin{equation}
            \hat{\mathbf{\Gamma}}_{\mathbf{X}} = \dfrac{1}{T}\mathbf{X}\mathbf{X}^\herm
        \end{equation}
        \State Perform the EVD of the sample covariance matrix:
        \begin{equation}
            \hat{\mathbf{\Gamma}}_{\mathbf{X}} = \tilde{\mathbf{U}}\tilde{\mathbf{\Lambda}}\tilde{\mathbf{U}}^\herm
        \end{equation}
        \State Order the eigenvectors by decreasing eigenvalues amplitude, and perform subspace separation: $\tilde{\mathbf{U}} = \left[\tilde{\mathbf{U}}_S, \tilde{\mathbf{U}}_N\right]$
        \State Compute the MUSIC spectrum along the angular grid:
        \begin{equation}\label{eq:music_grid}
            P_{\textrm{MUSIC}}\left(\boldsymbol{\theta}_g \vert \boldsymbol{\zeta}\right) = \dfrac{1}{\norm{\tilde{\mathbf{U}}_N^\herm \mathbf{A}_{\boldsymbol{\zeta}}\left(\boldsymbol{\theta}_g\right)}{2}^2}
        \end{equation}
        \State Estimate the DoAs as the argument of the $M$ top-peaks:
        \begin{equation}\label{eq:argmax_computation}
            \hat{\boldsymbol{\theta}} = \argmax_{\boldsymbol{\theta}_g,M} P_{\textrm{MUSIC}}\left(\boldsymbol{\theta}_g \vert \boldsymbol{\zeta}\right)
        \end{equation}
        \Ensure Estimated DoAs: $\hat{\boldsymbol{\theta}} \in \mathbb{R}^M$
    \end{algorithmic}
\end{algorithm}

\section{Proposed method}\label{sec:proposed_method}
As previously mentioned, the performance of the classical MUSIC algorithm deteriorates when considering hardware impairments. To address this issue, it is proposed to modify its structure to make it differentiable, enabling the learning of the array parametrization within the DoA estimation scheme.

\noindent\textbf{Model-based differentiable architecture.} Stochastic gradient descent (SGD) is leveraged to solve:
\begin{mini!}
    {\boldsymbol{\zeta}}{\espo{\mathcal{L}\left(\boldsymbol{\theta}, \hat{\boldsymbol{\theta}}\left(\mathbf{X}\vert\boldsymbol{\zeta}\right)\right)}{\left(\boldsymbol{\theta},\mathbf{X}\right) \sim \mathcal{P}_{\left(\boldsymbol{\theta},\mathbf{X}\right)}}, \tag{$\mathrm{P}1$}\label{eq:optim_problem}}
    {}{}
    \addConstraint{\boldsymbol{\zeta} \in \mathbb{C}^N\times \mathbb{R}^N}{\nonumber}
\end{mini!}
where $\mathcal{P}_{\left(\boldsymbol{\theta},\mathbf{X}\right)}$ represents the data distribution, $\mathcal{L}$ is a loss function on the DoAs, $\boldsymbol{\theta}$ are the true DoAs associated to the measurements $\mathbf{X}$ and $\hat{\boldsymbol{\theta}}\left(\mathbf{X}\vert\boldsymbol{\zeta}\right)$ are estimated DoAs from a given DoA estimator with array parametrization knowledge $\boldsymbol{\zeta}$. In order to solve~\eqref{eq:optim_problem} through SGD, it is necessary to compute the derivatives $\nabla_{\boldsymbol{\zeta}}\mathcal{L}(\boldsymbol{\theta}, \hat{\boldsymbol{\theta}}\left(\mathbf{X}\vert\boldsymbol{\zeta}\right))$. Using the derivative chain rule, it can easily be shown that one has to compute $\nabla_{\boldsymbol{\zeta}} \hat{\boldsymbol{\theta}}\left(\mathbf{X}\vert\boldsymbol{\zeta}\right)$ to compute $\nabla_{\boldsymbol{\zeta}}\mathcal{L}(\boldsymbol{\theta}, \hat{\boldsymbol{\theta}}\left(\mathbf{X}\vert\boldsymbol{\zeta}\right))$. 


\addtocounter{equation}{-1}

The MUSIC algorithm is inherently non-differentiable, a characteristic arising from the $\argmax$ operation in Eq.~\eqref{eq:argmax_computation}. This hard decision scheme can be represented through angular masks containing a single value at each spectrum peak, with no operations performed within those windows. Since only the operations following the applications of the fixed angular masks are considered for gradient computations, it leads to the non-existence of $\nabla_{\boldsymbol{\zeta}} \hat{\boldsymbol{\theta}}\left(\mathbf{X}\vert\boldsymbol{\zeta}\right)$ for the MUSIC method, which results in its non-differentiability.

It is thus proposed to replace this hard-decision scheme with a differentiable softmax-based approach outlined in Algorithm~\ref{alg:diffMUSIC}, where each estimated DoA can be seen as a convex combination of grid DoAs around the associated MUSIC spectrum peak. Note that $\boldsymbol{\theta}^{\mathrm{mask}}_i=\Pi_L(\boldsymbol{\theta}_g,\theta_{i}^{\mathrm{peak}})$ denotes a windowing operation of size $L$, applied on the DoA grid $\boldsymbol{\theta}_g$, centered around $\theta_{i}^{\mathrm{peak}}$ such that $\boldsymbol{\theta}^{\mathrm{mask}}_i \in \mathbb{R}^L$. As for MUSIC, the diffMUSIC method initially estimates the spectrum peaks using a non-differentiable peak finding method: $\argmax$ in Eq.~\eqref{eq:peak_finding_diffMUSIC}. Then, angular windows are defined around each peaks. As depicted in Eq.~\eqref{eq:diff_DoA_est}, within each window, the estimated DoAs are computed using differentiable operations involving the steering matrix $\mathbf{A}_{\boldsymbol{\zeta}}\left(\boldsymbol{\theta}_{g}\right)$, through the computed MUSIC spectrum. This ensures that $\nabla_{\boldsymbol{\zeta}} \hat{\boldsymbol{\theta}}\left(\mathbf{X}\vert\boldsymbol{\zeta}\right)$ exists for a fixed angular mask, thereby making the method differentiable. Note that, in addition to being differentiable, this approach also increases angular resolution as off-grid estimates are possible through the convex DoA combination.

\begin{algorithm}[tb]
    \caption{diffMUSIC algorithm.}
    \label{alg:diffMUSIC}
    \begin{algorithmic}[1]
        \Require Computed MUSIC spectrum $P_{\textrm{MUSIC}}\left(\boldsymbol{\theta}_g \vert \boldsymbol{\zeta}\right)$ from Eq.~\eqref{eq:music_grid}, angular window size $L$, current array parametrization knowledge $\boldsymbol{\zeta}$.
        \State Find peaks in $P_{\textrm{MUSIC}}\left(\boldsymbol{\theta}_g \vert \boldsymbol{\zeta}\right)$: $\boldsymbol{\theta}^{\mathrm{peak}} = \{\theta_{i}^{\mathrm{peak}}\}_{i=1}^M$:
        \begin{equation}\label{eq:peak_finding_diffMUSIC}
            \boldsymbol{\theta}^{\mathrm{peak}} = \argmax_{\boldsymbol{\theta}_g,M} P_{\textrm{MUSIC}}\left(\boldsymbol{\theta}_g \vert \boldsymbol{\zeta}\right)
        \end{equation}
        \ForEach{$\theta_{i}^{\mathrm{peak}}$}
            \State Compute an angular mask centered around $\theta_{i}^{\mathrm{peak}}$: 
            \begin{equation}\label{eq:angular_mask}
                \boldsymbol{\theta}^{\mathrm{mask}}_i = \Pi_L(\boldsymbol{\theta}_g,\theta_{i}^{\mathrm{peak}})
            \end{equation}
            \State Estimate the associated DoA:
            \begin{equation}\label{eq:diff_DoA_est}
                \hat{\theta}_i = \left(\boldsymbol{\theta}^{\mathrm{mask}}_i\right)^\transp \texttt{softmax}\left(P_{\mathrm{MUSIC}}\left(\boldsymbol{\theta}^{\mathrm{mask}}_i \vert \boldsymbol{\zeta}\right)\right)
            \end{equation}
        \EndFor
        \Ensure Estimated DoAs: $\hat{\boldsymbol{\theta}} \in \mathbb{R}^M$
    \end{algorithmic}
\end{algorithm}

\noindent\textbf{Learning framework.} As presented in~\cite{Shmuel2023a,Shmuel2023}, the root mean squared periodic error (RMSPE) on the DoAs can be used as a loss function to solve~\eqref{eq:optim_problem} in a supervised learning (SL) manner. This loss function is defined as:
\begin{equation}\label{eq:RMSPE_loss}
    \mathcal{L}_{\mathrm{SL},\theta} = \dfrac{1}{\abs{\mathcal{T}}} \sum_{\left(\boldsymbol{\theta},\mathbf{X}\right) \in \mathcal{T}} \min_{\mathbf{P} \in \mathcal{P}} \frac{\norm{\mathrm{mod}_{\pi} \left(\boldsymbol{\theta}-\mathbf{P}\hat{\boldsymbol{\theta}}\left(\mathbf{X}\vert \boldsymbol{\zeta}\right)\right)}{2}}{\sqrt{M}},
\end{equation}
where $\mathcal{P}$ is the set of permutation matrices, and $\mathcal{T} = \left\{\boldsymbol{\theta}_i, \mathbf{X}_i\right\}_{i=1}^{N_t}$ is the training set. The RMSPE takes into account the permutation invariance in the DoA learning task through the permutation matrices $\mathbf{P} \in \mathcal{P}$: more details can be found in~\cite{Shmuel2023a}. Rather than learning on estimated DoAs, a proposed alternative strategy is to maximize the MUSIC spectrum amplitude at the true DoA locations. The associated supervised loss is defined as:
\begin{equation}
    \mathcal{L}_{\mathrm{SL},P} = -\dfrac{1}{\abs{\mathcal{T}}} \sum_{\left(\boldsymbol{\theta},\mathbf{X}\right) \in \mathcal{T}}\sum_i P_{\mathrm{MUSIC}}\left(\theta_i \vert \boldsymbol{\zeta} \right).
\end{equation}
This approach does not directly solve~\eqref{eq:optim_problem}; instead, it addresses a proxy problem, the solution of which translates into minimizing the DoA estimation error. Indeed, minimizing $\mathcal{L}_{\mathrm{SL},P}$ can be interpreted as finding an array parametrization that maximizes the orthogonality between steering vectors associated to the true DoAs and the noise subspace.

Once the impairments are learned using $\mathcal{L}_{\mathrm{SL},P}$, the DoAs can then be estimated using the classical MUSIC algorithm, or through diffMUSIC, with the learned array. This alternative SL strategy presents a reduced computational complexity during training in comparison to the RMSPE based strategy. This is presented in Table~\ref{table:time_complexity}, where $\kappa_{\mathrm{peak}}$, resp. $\kappa_{\mathrm{EVD}}$, represents the time complexity associated to the peak finding function, resp. EVD. This complexity reduction arises from the fact that, with $\mathcal{L}_{\mathrm{SL},P}$, the MUSIC spectrum only needs to be evaluated at the true DoA locations, rather than over a fine angular grid, as required for computing the DoAs with $\mathcal{L}_{\mathrm{SL},\theta}$.

\noindent\textbf{Unsupervised learning.} The proposed SL approaches rely on true DoA knowledge, which may limit their practical applicability. An unsupervised learning (UL) strategy can be implemented by using a loss function designed to maximize the MUSIC spectrum peaks within each angular mask of diffMUSIC. Such loss function can for instance, be based on the coefficient of variation, the Gini coefficient, the kurtosis or the Kullback-Leibler divergence relative to the uniform distribution. It is proposed to consider a Jain's index (JI) based loss function where minimization translates to a sharp peak inside the masked spectrum. The JI is defined as:
\begin{equation}
    \forall \mathbf{x}\in \mathbb{R}^n, \text{ } \mathfrak{J}\left(\mathbf{x}\right) = \dfrac{\left(\sum_i x_i\right)^2}{n\sum_i x_i^2},
\end{equation}
and the associated unsupervised loss function is defined as:
\begin{equation}
    \mathcal{L}_{\mathrm{UL}} = \dfrac{1}{\abs{\mathcal{T}}} \sum_{\mathbf{X} \in \mathcal{T}} \sum_i \mathfrak{J}\left(P_{\mathrm{MUSIC}}\left(\boldsymbol{\theta}^{\mathrm{mask}}_i\left(\mathbf{X} \vert \boldsymbol{\zeta}\right)\vert \boldsymbol{\zeta}\right)\right),
\end{equation}
where $\boldsymbol{\theta}^{\mathrm{mask}}_i\left(\mathbf{X}\vert \boldsymbol{\zeta}\right)$ is the obtained angular mask for given measurements in Eq.~\eqref{eq:angular_mask}.

\begin{table}[t]
    \centering
    \begin{tabular}{cccc}
        \toprule
        & $\mathcal{L}_{\mathrm{SL},\theta}$ & $\mathcal{L}_{\mathrm{SL},P}$\\
        \midrule
        \makecell{Time\\ complexity} & $\mathcal{O}\left(N N_{\theta} + \kappa_{\mathrm{peak}} + \kappa_{\mathrm{EVD}}\right)$ & $\mathcal{O}\left(N M + \kappa_{\mathrm{EVD}}\right)$\\
        \bottomrule
    \end{tabular}
    \caption{Training complexity comparison: $M \ll N_{\theta}$}
    \label{table:time_complexity}
\end{table}

\begin{table*}[t]
    \centering
    \begin{tabular}{cccccccccccc}
        \toprule
        & & \multicolumn{4}{c}{Baselines} &  \multicolumn{2}{c}{$\mathcal{L}_{\mathrm{SL},\theta}$} &  \multicolumn{2}{c}{$\mathcal{L}_{\mathrm{SL},P}$} &  \multicolumn{2}{c}{$\mathcal{L}_{\mathrm{UL}}$} \\
        \cmidrule(lr){3-6} \cmidrule(lr){7-8} \cmidrule(lr){9-10} \cmidrule(lr){11-12} & & $\texttt{M}$ (nom.) & $\texttt{M}$ (phys.) & $\texttt{dM}$ (phys.) & \texttt{SubspaceNet} & $\texttt{M}$ & $\texttt{dM}$  & $\texttt{M}$ & $\texttt{dM}$ & $\texttt{M}$ & $\texttt{dM}$ \\
        \midrule
        \multirow[c]{2}{*}{\vspace{-2mm} RMSPE $\left(^\circ\right)$} & $M=1$ & $2.425$ & $0.014$ & $0.013$ & $0.098$ & $0.019$ & $0.015$ & $0.013$ & $\mathbf{0.013}$ & $1.339$ & $1.310$\\
        \cmidrule(l){2-12}
        & $M=5$ & $9.976$ & $4.358$ & $4.275$ & $16.123$ & $5.371$ & $5.178$ & $4.325$ & $\mathbf{4.209}$ & $4.834$ & $4.731$\\
        \bottomrule
    \end{tabular}
    \caption{Baseline comparisons}
    \label{table:perf_baselines}
\end{table*}

\begin{figure*}[htbp]
    \centering
    \begin{minipage}[b]{\columnwidth}
        \centering
        \centering
        \includegraphics[width=\columnwidth]{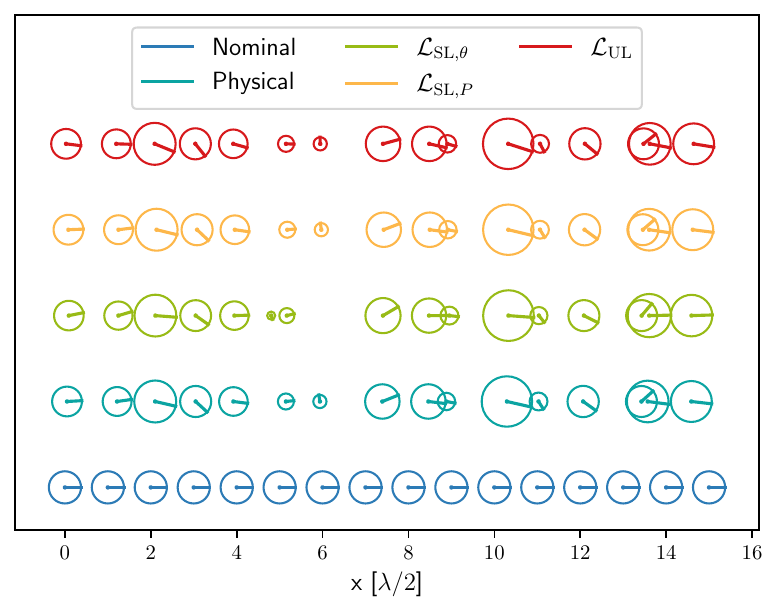}
        \caption{Learned parameters comparison for $M=5$, $30$dB SNR}
        \label{fig:learned_params}
    \end{minipage}
    \hfill
    \begin{minipage}[b]{\columnwidth}
        \centering
        \includegraphics[width=\columnwidth]{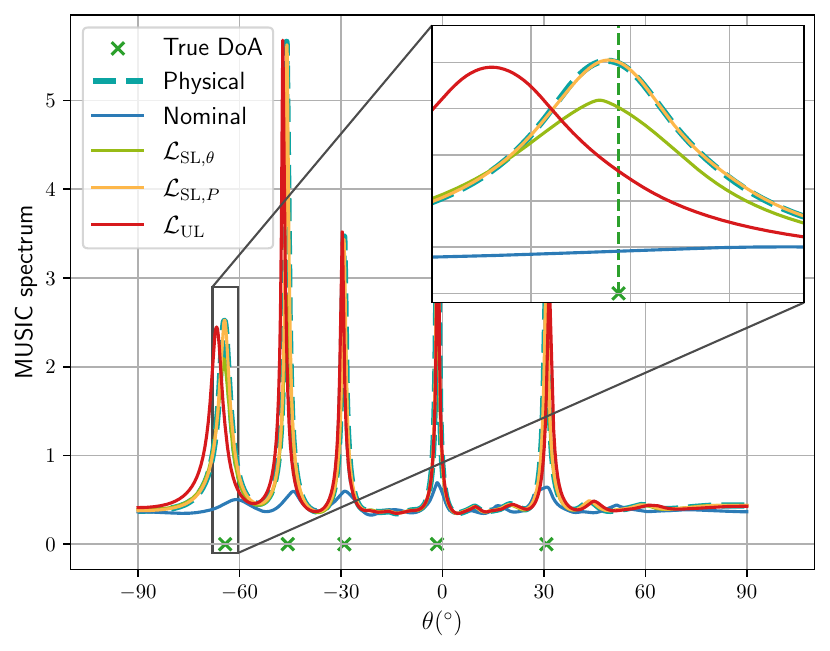}
        \caption{MUSIC spectrums with learned arrays, $M=5$, $T=100$, $10$dB SNR}
        \label{fig:learned_spectrums}
    \end{minipage}
\end{figure*}

\begin{figure*}[htbp]
    \centering
    \begin{minipage}[b]{\columnwidth}
        \centering
        \includegraphics[width=\columnwidth]{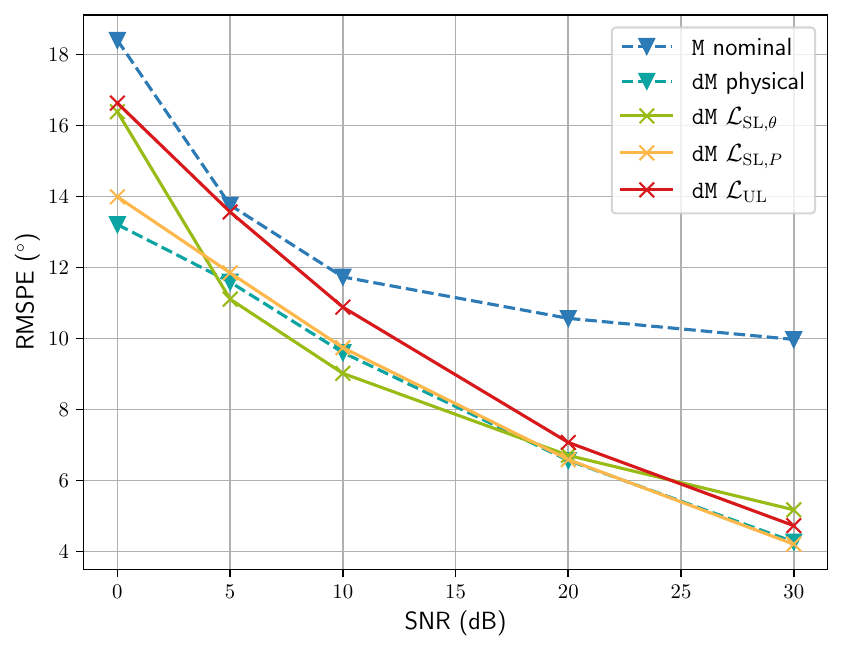}
        \caption{RMSPE evolution with sensing SNR, $T=100$}
        \label{figs:RMSPE_vs_SNR}
    \end{minipage}
    \hfill
    \begin{minipage}[b]{\columnwidth}
        \centering
        \includegraphics[width=\columnwidth]{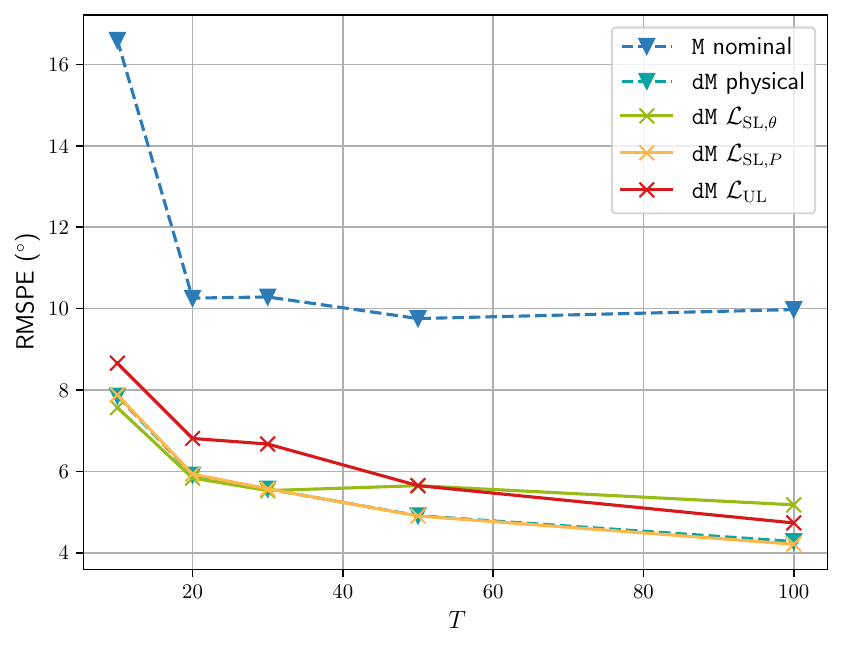}
        \caption{RMSPE evolution with snapshot number, $30$dB SNR}
        \label{figs:RMSPE_vs_T}
    \end{minipage}
\end{figure*}

\section{Experiments}\label{sec:Experiments}
It is proposed to study the impairments learning performance of the proposed diffMUSIC method against several baselines and in different scenarios.

\noindent\textbf{Dataset generation.} A ULA with $N=16$ antennas is considered. Measurements are generated following Eq.~\eqref{eq:system_model}. The sources DoAs are generated as $\forall i \in \llbracket 1,M\rrbracket, \theta_i \sim \mathcal{U}\left[-80^\circ, 80^\circ\right]$. The sensing signal to noise ratio (SNR) is computed as $\mathrm{SNR} = 10\log_{10}\left(\sigma^2_s/\sigma^2_n\right)$. Concerning the hardware impairments, imperfections in both antenna locations and gains are taken into account. For the antenna locations, one has:
\begin{equation}
    \forall i \in \llbracket 1, M\rrbracket, p_i = \tilde{p}_i + \delta_{p_i},
\end{equation}
where $p_i$, resp. $\tilde{p}_i$, denotes the physical, resp. nominal, $i$th antenna location, and $\delta_{p_i} \sim \mathcal{U}\left[-\eta, \eta\right]$, with $\eta \propto \lambda/2$. Note that the locations are only altered along the non-null dimension of the ULA. Similarly, for the antenna gains, one has:
\begin{equation}
    \forall i \in \llbracket 1, M \rrbracket, g_i = \tilde{g}_i + \delta_{g_i},
\end{equation}
where $g_i$, resp. $\tilde{g}_i$, denotes the physical, resp. nominal, complex gain for the $i$th antenna, and $\delta_{g_i} \sim \mathcal{CN}\left(0,\sigma^2_g\right)$.

Unless otherwise stated, $M=5$ sources, $T=100$ snapshots, $\eta=0.5 \lambda/2$ and $\sigma^2_{g} = 0.36$ are considered.

\noindent\textbf{Baselines.} The proposed approach is compared against the classical MUSIC algorithm (denoted as $\texttt{M}$) with nominal and physical array, but also against diffMUSIC (denoted as $\texttt{dM}$) with physical array. Note that the window size $L$ of diffMUSIC is optimized in each considered scenario. Additionally, it is proposed to compare diffMUSIC against the SubspaceNet network presented in~\cite{Shmuel2023}: as its goal is to learn a surrogate covariance matrix to counter imperfections in the sensing process, it is able to take into account hardware impairments. The evaluation metric for all approaches is the RMSPE defined in Eq.~\eqref{eq:RMSPE_loss} evaluated over a test set.

\noindent\textbf{Performance against baselines.} Table~\ref{table:perf_baselines} presents a DoA estimation performance comparison for $M=\left\{1,5\right\}$ sources, at $30$dB SNR. It can be seen that the proposed approaches outperform the MUSIC algorithm with nominal array knowledge. Moreover, one can see that SubspaceNet performs relatively well for $M=1$ but totally fails to learn the imperfections in the multi-source scenario. Additionally, it can be seen that diffMUSIC with $\mathcal{L}_{\mathrm{SL},P}$ equals or outperforms diffMUSIC with physical array, indicating good impairment learning capabilities. Furthermore, one can remark that the UL approach presents good performance in the multi-sources scenario: this is of particular interest as this approach does not rely on labeled DoAs. Note that this approach could be extended in an online learning strategy where the antenna array would learn its hardware impairments on the fly.

\noindent\textbf{Learned parameters and MUSIC spectrums.} It is proposed to visualize the learned parameters by the proposed methods for $M=5$ and their impact on the associated MUSIC spectrums. Fig.~\ref{fig:learned_params} presents the learned parameters: the center of each circle represents an antenna location, the circle radius represents the associated complex gain magnitude, and the segment angle represents the associated complex gain phase. It can be seen that the $\mathcal{L}_{\mathrm{SL},\theta}$ approach presents good impairments learning except for two antennas whose defaults are small. It is worth noting that both the $\mathcal{L}_{\mathrm{SL},P}$ and $\mathcal{L}_{\mathrm{UL}}$ approaches achieve near-perfect learning of the impairments which aligns with their strong performance observed in Table~\ref{table:perf_baselines}.

Fig.~\ref{fig:learned_spectrums} presents the spectrums obtained with the learned parameters by the different approaches. Results are illustrated at $10$dB SNR for better visualization. One can remark that, as expected by its definition, the spectrum obtained with the array parametrization learned through $\mathcal{L}_{\mathrm{SL},P}$ tightly follows the physical spectrum. One can also note that the peaks of the spectrum obtained through $\mathcal{L}_{\mathrm{SL},\theta}$ are aligned with the ones of the physical spectrum. Unsurprisingly, an angular shift is observed in the peaks of the spectrum obtained through $\mathcal{L}_{\mathrm{UL}}$: while this approach allows to maximize peak amplitudes, its unsupervised nature does not allow to add angular bias on the peak locations during training. This angular shift phenomenon is particularly noticeable near the endfire of the antenna array.

\noindent\textbf{Performance against sensing noise.} Fig.~\ref{figs:RMSPE_vs_SNR} presents the DoA estimation performance evolution with variable SNR, for $M=5$ sources and $T=100$ snapshots. One can remark that, while the $\mathcal{L}_{\mathrm{SL},\theta}$ approach presents good performance in the high SNR region, the $\mathcal{L}_{\mathrm{SL},P}$ approach presents the overall best performance as it follows tightly the diffMUSIC performance with physical array. It is worth noting that, in the very low SNR region, both the $\mathcal{L}_{\mathrm{SL},\theta}$ and $\mathcal{L}_{\mathrm{UL}}$ approaches are significantly affected by noise, preventing them from effectively learning hardware impairments. In contrast, the $\mathcal{L}_{\mathrm{SL},P}$ approach does not suffer from that issue.

\noindent\textbf{Performance against snapshot number.}  Fig.~\ref{figs:RMSPE_vs_T} presents the DoA estimation performance evolution with variable number of snapshots for $M=5$ sources and $30$dB SNR. As expected, the estimation performance degrades across all approaches when the number of snapshots is low. This can be attributed to the higher estimation error in the empirical covariance matrix when $T$ is low. It is also worth noting that the $\mathcal{L}_{\mathrm{SL},P}$ approach exhibits the overall best performance. It demonstrates its superiority against other methods as it offers both better estimation performance in low SNR or low snapshot scenarios, while also achieving lower time complexity.

\section{Conclusion and future work}\label{sec:conclusions}
This paper discussed the application of a differentiable version of the MUSIC algorithm for learning hardware impairments in the DoA estimation context. The differentiable MUSIC algorithm was obtained by replacing the non-differentiable argmax search by a softmax-based approach, where each estimated DoA can be viewed as convex combination of DoAs. Additionally, both supervised and unsupervised loss functions have been proposed for this impairment learning task. The proposed methods have been evaluated against several baselines, showing their good hardware impairment learning and DoA estimation abilities. Specifically, it has been shown that the proposed methods are able to learn hardware impairments even in the presence of high sensing noise and low number of measurement snapshots. Furthermore, the unsupervised learning strategy is of particular interest as it enables the antenna system to perform hardware impairment compensation without requiring measured DoAs.

Future work will focus on extending the differentiable MUSIC approach to address coherent or partially correlated sources, and its application to near-field localization.

\bibliographystyle{IEEEtran}
\bibliography{references}

\begin{thebibliography}{10}
\providecommand{\url}[1]{#1}
\csname url@samestyle\endcsname
\providecommand{\newblock}{\relax}
\providecommand{\bibinfo}[2]{#2}
\providecommand{\BIBentrySTDinterwordspacing}{\spaceskip=0pt\relax}
\providecommand{\BIBentryALTinterwordstretchfactor}{4}
\providecommand{\BIBentryALTinterwordspacing}{\spaceskip=\fontdimen2\font plus
\BIBentryALTinterwordstretchfactor\fontdimen3\font minus
  \fontdimen4\font\relax}
\providecommand{\BIBforeignlanguage}[2]{{%
\expandafter\ifx\csname l@#1\endcsname\relax
\typeout{** WARNING: IEEEtran.bst: No hyphenation pattern has been}%
\typeout{** loaded for the language `#1'. Using the pattern for}%
\typeout{** the default language instead.}%
\else
\language=\csname l@#1\endcsname
\fi
#2}}
\providecommand{\BIBdecl}{\relax}
\BIBdecl

\bibitem{pesavento2023three}
M.~Pesavento, M.~Trinh-Hoang, and M.~Viberg, ``Three more decades in array
  signal processing research: An optimization and structure exploitation
  perspective,'' \emph{IEEE Signal Process. Mag.}, vol.~40, no.~4, pp. 92--106,
  2023.

\bibitem{Corlay24}
V.~Corlay, V.-H. Nguyen, and N.~Gresset, ``Probabilistic positioning via ray
  tracing with noisy angle of arrival measurements,'' \emph{IEEE Commun.
  Lett.}, vol.~28, no.~10, pp. 2278--2282, 2024.

\bibitem{Schmidt1986}
R.~Schmidt, ``Multiple emitter location and signal parameter estimation,''
  \emph{IEEE Trans. on Antennas and Propag.}, vol.~34, no.~3, pp. 276--280,
  Mar. 1986.

\bibitem{Barabell83}
A.~Barabell, ``Improving the resolution performance of eigenstructure-based
  direction-finding algorithms,'' in \emph{IEEE Int. Conf. on Acoust., Speech,
  and Signal Process.}, vol.~8, 1983, pp. 336--339.

\bibitem{VanVeen88}
B.~Van~Veen and K.~Buckley, ``Beamforming: a versatile approach to spatial
  filtering,'' \emph{IEEE ASSP Mag.}, vol.~5, no.~2, pp. 4--24, 1988.

\bibitem{Roy1989}
R.~Roy and T.~Kailath, ``Esprit-estimation of signal parameters via rotational
  invariance techniques,'' \emph{IEEE Trans. on Acoust., Speech, and Signal
  Process.}, vol.~37, no.~7, pp. 984--995, Jul. 1989.

\bibitem{Vasylyshyn09}
V.~Vasylyshyn, ``Direction of arrival estimation using esprit with sparse
  arrays,'' in \emph{European Radar Conf.}, 2009, pp. 246--249.

\bibitem{Björnson13}
E.~Björnson, P.~Zetterberg, M.~Bengtsson, and B.~Ottersten, ``Capacity limits
  and multiplexing gains of mimo channels with transceiver impairments,''
  \emph{IEEE Commun. Lett.}, vol.~17, no.~1, pp. 91--94, 2013.

\bibitem{Björnson14}
E.~Björnson, J.~Hoydis, M.~Kountouris, and M.~Debbah, ``Massive mimo systems
  with non-ideal hardware: Energy efficiency, estimation, and capacity
  limits,'' \emph{IEEE Trans. on Inf. Theory}, vol.~60, no.~11, pp. 7112--7139,
  2014.

\bibitem{Björnson15}
E.~Björnson, M.~Matthaiou, A.~Pitarokoilis, and E.~G. Larsson, ``Distributed
  massive mimo in cellular networks: Impact of imperfect hardware and number of
  oscillators,'' in \emph{2015 23rd European Signal Process. Conf.}, 2015, pp.
  2436--2440.

\bibitem{Höhne10}
T.~Höhne and V.~Ranki, ``Phase noise in beamforming,'' \emph{IEEE Trans. on
  Wireless Commun.}, vol.~9, no.~12, pp. 3682--3689, 2010.

\bibitem{chatelier_impact_2022}
B.~Chatelier and M.~Crussière, ``On the {Impact} of {Phase} {Noise} on
  {Beamforming} {Performance} for {mmWave} {Massive} {MIMO} {Systems},'' in
  \emph{2022 {IEEE} {Wireless} {Commun.} and {Netw.} {Conf.}}, Apr. 2022, pp.
  1563--1568.

\bibitem{Bozorgi21}
F.~Bozorgi, P.~Sen, A.~N. Barreto, and G.~Fettweis, ``{RF} front-end challenges
  for joint communication and radar sensing,'' in \emph{IEEE Int. Online Symp.
  on Joint Commun. \& Sens.}, 2021, pp. 1--6.

\bibitem{Tubail24}
D.~A. Tubail and S.~S. Ikki, ``Bayesian {Cramer-Rao} bound, extended and
  unscented {Kalman} filters based tracking through non-ideal transceivers in
  {5G} and beyond,'' \emph{IEEE Trans. on Veh. Technol.}, pp. 1--13, 2024.

\bibitem{Chen24}
H.~Chen, M.~F. Keskin, S.~R. Aghdam, H.~Kim, S.~Lindberg, A.~Wolfgang, T.~E.
  Abrudan, T.~Eriksson, and H.~Wymeersch, ``Modeling and analysis of
  {OFDM}-based {5G/6G} localization under hardware impairments,'' \emph{IEEE
  Trans. on Wireless Commun.}, vol.~23, no.~7, pp. 7319--7333, 2024.

\bibitem{OShea2017}
T.~O’Shea and J.~Hoydis, ``An introduction to deep learning for the physical
  layer,'' \emph{IEEE Trans. Cogn. Commun. Netw.}, vol.~3, no.~4, pp. 563--575,
  2017.

\bibitem{Wang2017}
T.~Wang, C.-K. Wen, H.~Wang, F.~Gao, T.~Jiang, and S.~Jin, ``Deep learning for
  wireless physical layer: Opportunities and challenges,'' \emph{China
  Commun.}, vol.~14, no.~11, pp. 92--111, 2017.

\bibitem{Alkhateeb2018}
A.~Alkhateeb, S.~Alex, P.~Varkey, Y.~Li, Q.~Qu, and D.~Tujkovic, ``Deep
  learning coordinated beamforming for highly-mobile millimeter wave systems,''
  \emph{IEEE Access}, vol.~6, pp. 37\,328--37\,348, 2018.

\bibitem{Balevi2020}
E.~Balevi, A.~Doshi, and J.~G. Andrews, ``Massive {MIMO} channel estimation
  with an untrained deep neural network,'' \emph{IEEE Trans. on Wireless
  Commun.}, vol.~19, no.~3, pp. 2079--2090, 2020.

\bibitem{Chatelier2023b}
B.~Chatelier, V.~Corlay, C.~Ciochina, F.~Coly, and J.~Guillet, ``Influence of
  dataset parameters on the performance of direct ue positioning via deep
  learning,'' in \emph{Joint European Conf. on Netw. and Commun./6G Summit},
  2023, pp. 126--131.

\bibitem{Corlay2022}
V.~Corlay, J.~J. Boutros, P.~Ciblat, and L.~Brunel, ``Neural network approaches
  to point lattice decoding,'' \emph{IEEE Trans. on Inf. Theory}, vol.~68,
  no.~5, pp. 2969--2989, 2022.

\bibitem{Shlezinger23}
N.~Shlezinger and Y.~C. Eldar, ``Model-based deep learning,'' \emph{Foundations
  and Trends® in Signal Process.}, vol.~17, no.~4, pp. 291--416, 2023.

\bibitem{Shlezinger23b}
N.~Shlezinger, J.~Whang, Y.~C. Eldar, and A.~G. Dimakis, ``Model-based deep
  learning,'' \emph{Proc. of the IEEE}, vol. 111, no.~5, pp. 465--499, 2023.

\bibitem{yassine2022}
T.~Yassine and L.~Le~Magoarou, ``{mpNet}: Variable depth unfolded neural
  network for massive {MIMO} channel estimation,'' \emph{IEEE Trans. on
  Wireless Commun.}, vol.~21, no.~7, pp. 5703--5714, 2022.

\bibitem{Chatelier2022efficient}
B.~Chatelier, L.~Le~Magoarou, and G.~Redieteab, ``Efficient deep unfolding for
  {SISO-OFDM} channel estimation,'' in \emph{IEEE Int. Conf. on Commun.}, 2023.

\bibitem{Lavi23}
O.~Lavi and N.~Shlezinger, ``Learn to rapidly and robustly optimize hybrid
  precoding,'' \emph{IEEE Trans. on Commun.}, vol.~71, no.~10, pp. 5814--5830,
  2023.

\bibitem{Yassine23}
T.~Yassine, B.~Chatelier, V.~Corlay, M.~Crussière, S.~Paquelet, O.~Tirkkonen,
  and L.~L. Magoarou, ``Model-based deep learning for beam on prediction based
  on a channel chart,'' in \emph{2023 57th Asilomar Conf. on Signals, Syst.,
  and Comput.}, 2023, pp. 1636--1640.

\bibitem{Samuel17}
N.~Samuel, T.~Diskin, and A.~Wiesel, ``Deep mimo detection,'' in \emph{2017
  IEEE 18th Int. Workshop on Signal Process. Advances in Wireless Commun.},
  2017, pp. 1--5.

\bibitem{mateos2023model}
J.~M. Mateos-Ramos, C.~H{\"a}ger, M.~F. Keskin, L.~Le~Magoarou, and
  H.~Wymeersch, ``Model-based end-to-end learning for multi-target integrated
  sensing and communication,'' \emph{arXiv preprint arXiv:2307.04111}, 2023.

\bibitem{Mateos_Ramos24}
J.~M. Mateos-Ramos, B.~Chatelier, C.~Häger, M.~F. Keskin, L.~Le~Magoarou, and
  H.~Wymeersch, ``Semi-supervised end-to-end learning for integrated sensing
  and communications,'' in \emph{IEEE Int. Conf. on Mach. Learn. for Commun.
  and Netw.}, 2024, pp. 132--138.

\bibitem{Wu2019}
L.~Wu, Z.-M. Liu, and Z.-T. Huang, ``Deep convolution network for direction of
  arrival estimation with sparse prior,'' \emph{IEEE Signal Process. Lett.},
  vol.~26, no.~11, pp. 1688--1692, Nov. 2019.

\bibitem{Cong2021}
J.~Cong, X.~Wang, M.~Huang, and L.~Wan, ``Robust {DoA} estimation method for
  mimo radar via deep neural networks,'' \emph{IEEE Sensors J.}, vol.~21,
  no.~6, pp. 7498--7507, Mar. 2021.

\bibitem{Papageorgiou2021}
G.~Papageorgiou, M.~Sellathurai, and Y.~Eldar, ``Deep networks for
  direction-of-arrival estimation in low snr,'' \emph{IEEE Trans. on Signal
  Process.}, vol.~69, pp. 3714--3729, 2021.

\bibitem{Lan2023}
X.~Lan, H.~Zhai, and Y.~Wang, ``A novel {DoA} estimation of closely spaced
  sources using attention mechanism with conformal arrays,'' \emph{IEEE
  Access}, vol.~11, pp. 44\,010--44\,018, 2023.

\bibitem{Ji2024}
J.~Ji, W.~Mao, F.~Xi, and S.~Chen, ``Transmusic: A transformer-aided subspace
  method for {DoA} estimation with low-resolution adcs,'' in \emph{IEEE Int.
  Conf. on Acoust., Speech and Signal Process.}\hskip 1em plus 0.5em minus
  0.4em\relax IEEE, Apr. 2024, pp. 8576--8580.

\bibitem{Shmuel2023a}
D.~H. Shmuel, J.~P. Merkofer, G.~Revach, R.~J.~G. van Sloun, and N.~Shlezinger,
  ``Deep root music algorithm for data-driven {DoA} estimation,'' in \emph{IEEE
  Int. Conf. on Acoust., Speech and Signal Process.}, vol.~34.\hskip 1em plus
  0.5em minus 0.4em\relax IEEE, Jun. 2023, pp. 1--5.

\bibitem{Shmuel2023}
------, ``Subspacenet: Deep learning-aided subspace methods for {DoA}
  estimation,'' \emph{arXiv preprint arXiv:2306.02271}, 2023.

\bibitem{Stoica2005}
P.~G. Stoica and R.~Moses, \emph{Spectral analysis of signals}, P.~Stoica and
  R.~L. Moses, Eds.\hskip 1em plus 0.5em minus 0.4em\relax Upper Saddle River,
  NJ: Pearson, Prentice Hall, 2005, includes bibliographical references (p.
  423-434) and index.

\end{thebibliography}

\end{document}